\documentclass{jltp}

\usepackage{amssymb}
\usepackage{bm}

\title{Maximal length of trapped 
one-dimensional \\ Bose-Einstein condensates}

\author{Uwe R. Fischer}

\address{Eberhard-Karls-Universit\"at T\"ubingen, 
Institut f\"ur Theoretische Physik \\
Auf der Morgenstelle 14, 
D-72076 T\"ubingen, Germany}

\runninghead{Uwe R. Fischer}{Maximal length of 
Bose-Einstein condensates}

\begin{document}

\maketitle

\begin{abstract}
I discuss  a Bogoliubov inequality 
for obtaining a rigorous bound on the maximal axial extension of 
inhomogeneous 
one-di\-mensional Bose-Einstein condensates. 
An explicit upper limit for the aspect ratio of a strongly elongated, 
harmonically trapped Thomas-Fermi condensate is derived. 

\end{abstract}

The question on the existence and dimension dependence of off-diagonal 
long-range order, for an infinitely extended 
system of scalar uncharged bo\-sons, was 
conclusively answered by Hohenberg\cite{Hohenberg}.
He made use of the fact that in the thermodynamic limit of an infinite system 
of {\em interacting} particles of {bare} mass $m$, the relation 
(we set $\hbar = k_{\rm B} =1$ in what follows, 
where $k_{\rm B}$ is the Boltzmann constant) 
\begin{equation}
N_{\bm k} \ge -\frac12 + \frac{m T}{k^2} \frac{n_0}n \label{Theorem}
\end{equation}
for the occupation numbers 
$N_{\bm k}=< \hat b_{\bm k}^\dagger \hat b_{\bm k} >$ 
of plane wave states enumerated by ${\bm k}$ ($\ne \bm 0$) 
holds, where $T$ is the temperature and $m$ the bare mass of the
{\em interacting} particles. 
Angled brackets indicate a thermal
 ensemble (quasi-)average\cite{Bogoliubov}; $n_0$ and
$n$ are the condensate density, associated to ${\bm k} =0$, 
and total density, respectively. 
The inequality (\ref{Theorem}) is essentially a particular representation of  
the 
Bogoliubov $1/k^2$ theorem on  
correlation functions\cite{Bogoliubov}. 

The {\em Hohenberg theorem} is based on the mathematical fact 
that inequality (\ref{Theorem}) leads to a 
contradiction in dimension $D\le 2$ due to the (infrared) 
divergence of the wave vector integral of (\ref{Theorem}), which 
determines the number density of non-condensate atoms. We denote  
the infrared wave vector cutoff by $k_c= 2\pi /R$, 
where $R$ is the size of the system, tending to infinity (hence 
$k_c \rightarrow 0$), and the 
ultraviolet cutoff (the microscopic discretization length) 
we denote $k_{\rm Pl}$. Taking 
the integral of both sides of (\ref{Theorem}), 
we obtain in the one-dimensional case, using $N= nR$  
\renewcommand{\arraystretch}{1.5}
\begin{eqnarray}
{N-N_0} & 
\stackrel{R\rightarrow \infty}{\ge} & 
\frac{\textstyle mT}{\textstyle 2\pi k_{\rm Pl}} \, n_0 \, R^2 .
\label{1D2DDivergence}
\renewcommand{\arraystretch}{1.0}
\end{eqnarray}
The number of excited atoms $N-N_0$, where $N_0$ 
is the number of particles in the condensate, has to be finite. 
Therefore, condensates
cannot exist at any finite temperature, 
because $n_0=0$ 
necessarily in the thermodynamic limit of $R\rightarrow \infty$, due to 
the divergence of the right-hand side.
Physically, long-range thermal fluctuations, the influence of which becomes
increasingly dominant the lower the dimension of the system, destroy the 
coherence and long-range order expressed
by the existence of the condensate.

That there is indeed a difference of trapped Bose-Einstein-condensed vapors 
of reduced dimensionality\cite{Ketterle1D2D,PhaseFluctuation,OttMicro,Hellweg,Moritz,Stoeferle} 
to the thermodynamics in an infinite box 
limit stems from various reasons. 
First of all, the classification of excited non-condensate
states by plane waves is not the suitable one in a trapped gas: 
The condensation, in the limit of large total particle number $N$,  
takes place  primarily 
into a single particle state in co-ordinate space\cite{MullinJLTP97},
and condensate and total densities have (in principle arbitrary) 
spatial dependence, $n_0\rightarrow n_0 ({\bm r})$, $n \rightarrow n({\bm
  r})$. Second, there arises the question how interaction  
influences the existence of the condensate, given that inhomogeneity of 
$n_0 ({\bm r})$ and $n({\bm  r})$. 
An interacting
gas generally has a behavior increasingly different from the ideal gas the 
lower the dimension of the system\cite{LiebLiniger,KTBBEC,Lieb2003}. 

For comparison of the long-range coherence properties of trapped 
Bose-Einstein condensates with the spatially homogeneous case of the 
Hohenberg theorem, it is desirable to obtain a relation which is analogous 
to (\ref{Theorem}) in as close as possible a manner\cite{HT}.
It is important to stress that a major strength of 
(\ref{Theorem}) is the explicit interaction independence; 
this holds true as long as this interaction is velocity 
independent, i.e., depends only on the (relative) position 
of the particles\cite{Hohenberg,PinesNozieres}. 
One primary goal thus is to maintain this 
independence of the inequality on particle interactions.

Like Hohenberg in his original paper\cite{Hohenberg}, 
I start my analysis from the Bogoliubov inequality 
\cite{Bogoliubov}, which reads 
\begin{equation}
\frac12\left< \left\{ \hat A, \hat A^\dagger \right\} \right>\ge 
\frac{T \left|\left<[\hat C, \hat A] \right> \right|^2}
{\left<\left[\left[\hat C,\hat H\right],\hat C^\dagger\right] \right>} 
\label{Bogoliubov}
\end{equation}
for any two operators $\hat A$ and $\hat C$.
It is valid for any many-body quantum system, for which 
the thermal (quasi-)averages indicated by the angled brackets 
are well-defined and finite. 
This inequality was used by Mermin and Wagner to prove the absence
of ferro- and antiferromagnetism in one and two dimensions\cite{MerminWagner}, 
in a fashion analogous to Hohenberg's proof of the non-existence of 
off-diagonal long-range order for scalar particles in one and two spatial 
dimensions. 

The operators in relation (\ref{Bogoliubov}) are chosen 
to be the following {\em smeared} excitation and total density operators
(we keep the spatial dimension $D$ of the system for generality): 
\begin{eqnarray}
\hat A 
& \equiv & {\hat A}_{\bm k}=
\int_{{\cal D}_0} d^Dr 
\, \exp [i{\bm k}\cdot{\bm r}] 
\delta \hat \Phi ({\bm r}), 
\quad \;
\hat C 
=  \int_{{\cal D}_0} d^Dr \, \Phi^*_0 ({\bm r}) 
\hat \rho({\bm r}),
\label{fg(r)def}
\end{eqnarray}
where ${\cal D}_0$ is the domain of support of the condensate 
wave function $\Phi_0({\bm r})$, i.e., its (suitably chosen) 
normalization volume.
The excitation operator is defined by  
$\delta \hat \Phi ({\bm r}) = \hat \Phi ({\bm r})- \hat b_0 
\Phi_0({\bm r}) 
$ and the condensate wave function $\Phi_0({\bm r})$ 
is normalized to unity.
This choice of operators is analogous 
to the original Hohenberg theorem derivation in\cite{Hohenberg}, 
where instead of the position space basis chosen here, a
plane wave basis appropriate for the thermodynamically infinite system 
in a box was taken. In addition, I introduce a {\em variational}
wave number ${\bm k}$, which is chosen subject to the constraint that 
the right-hand side of (\ref{Bogoliubov}) is maximized (see the example
below). 
In position space, the smearing procedure 
is necessary, because 
products of two quantum field operators diverge if 
taken at  the same point in space.
An important step in the derivation of a position space version of Hohenberg's
theorem is to recognize that the so-called mixed response 
(the ``anomalous'' commutator) is taking the nonlocal form
\begin{eqnarray}
\left< \left[\hat \rho ({\bm r}), \delta\hat\Phi
({\bm r}')\right]\right>  
= 
{\sqrt{N_0}}\,\Phi_0({\bm r})
\left[ -
\delta ( {\bm r}- {\bm r}')
+\Phi^*_0({\bm r}) \Phi_0 ({\bm r}')\right]. 
\label{Phi0Commutator}
\end{eqnarray}
Here, {\em after} carrying out the commutator, 
$<\hat b_0 >=\sqrt{N_0}= <\hat b^\dagger_0>$ 
has been used, where this assignment is valid to 
$O(1/\sqrt{N_0})$.
The difference to the usual Bogoliubov treatment, 
which would neglect the second term on the right-hand side
of the canonical commutator $[ \delta \hat\Phi ({\bm r}), \delta 
\hat \Phi^\dagger ({\bm r}') ] =  
\delta ( {\bm r}- {\bm r}') - \Phi_0({\bm r}) \Phi^*_0({\bm r}')$
altogether, is that the second term in the mixed response function above 
would disappear in the Bogoliubov approximation.  
I will argue below that neglect of this term in the mixed response, 
for finite systems, would lead to a contradiction with existing, 
i.e., experimentally realized Bose-Einstein condensates. 
The Bogoliubov approach of $[\hat b_0, \hat b^\dagger_0]\equiv 0$ 
violates particle number conservation to $O(f)$, 
where $f=\sqrt{1-N_0/N}$, while the present approach resulting in the
mixed response (\ref{Phi0Commutator}), 
violates it only to $O(f^2)$, and this difference, 
which is quite significant for sufficiently large systems, 
proves indeed crucial for condensates to exist\cite{CastinDum}.
Using the above definitions and a position space
version of the $f$-sum rule\cite{PinesNozieres}, 
to eliminate details of the Hamiltonian and thus the interaction  
occurring in the denominator on the right-hand side of 
(\ref{Bogoliubov}), the final Bogoliubov inequality takes the form\cite{HT}:  
\begin{equation}
\frac{1-F}F \ge \frac1N\frac{2\pi R_c^2}{\lambda_{\rm dB}^2} 
\,{\cal C} ({\bm k}) 
-\frac{1}{2N_0}\left(1 - 
|\tilde \Phi_0 ({\bm k})|^2/\Omega_0\right).
\label{RcCF}
\end{equation}
Here, the condensate fraction $F= N_0/N$ and 
the de Broglie thermal wavelength $\lambda_{\rm dB} = \sqrt{2\pi /mT}$.
I define the 
{\em effective radius of curvature} of the condensate to be the 
curvature radius of the condensate wave function, weighted
by the total density distribution:  
\begin{equation}
{R_c}\equiv \sqrt{\frac{N}{\Omega_0}}\! \left(
{\int_{{\cal D}_0}\!\! d^Dr\,  \Phi_0({\bm r})\left[
-\Delta_{{\bm r}}\Phi_0^*({\bm r})\right] n ({\bm r})}\!
\right)^{-1/2}\!. \label{Rcdef}
\end{equation}
The variational functional ${\cal C} ({\bm k})$ is given by 
\begin{equation}
{\cal C} ({\bm k})= \left|\tilde n_0({\bm k}) -\tilde \Phi_0 ({\bm k})
\int_{{\cal D}_0} d^Dr\, \Phi_0^*({\bm r}) |\Phi_0({\bm r})|^2 
\right|^2, \label{Ckdef}
\end{equation} 
where the Fourier transforms of single particle 
condensate density and wave function are defined
to be $\tilde n_0({\bm k}) =\int d^Dr \,
|\Phi_0 ({\bm r})|^2 \exp[i{\bm k}\cdot {\bm r}]$ and 
$\tilde \Phi_0 ({\bm k}) =\int d^Dr\, 
\Phi_0 ({\bm r}) \exp[i{\bm k}\cdot {\bm r}]$.

There are various notable features of the inequality 
(\ref{RcCF}). First of all, it is seen that 
the two-dimensional trapped case is marginal: In the 2D 
case, the radius of curvature $R_c$ scales like 
$R_c \propto N^{-1/2}$. 
Indeed, in the {trapped} gas, the logarithmic 
divergence of the 2D case in (\ref{1D2DDivergence}) is 
cut off by the trapping potential, and Bose-Einstein 
condensates can exist even in a (suitably defined) 
thermodynamic limit\cite{LiebSeiringer}. Second, 
the value of ${\cal C} ({\bm k})$ is strongly 
reduced because of the second term under  the  square in (\ref{Ckdef}); 
this is a consequence of the fact that we do not make use of the Bogoliubov 
approximation and, therefore, the second term in the 
mixed response (\ref{Phi0Commutator}) is included. 
Third, the relation represents, after  the set 
$\{N_0, \Phi_0({\bm r}), n({\bm r}) \}$ (and thus also $F$) 
has been specified at a given 
temperature, an inequality representing primarily a geometric statement: 
It gives a bound on the possible ratio of mean curvature radius
$R_c$ (which is dominated by the weakly confining directions), and the 
de Broglie wavelength $\lambda_{\rm dB}$. I also mention here  
that the final term in (\ref{RcCF}), which corresponds to the
``$-1/2$'' on the right-hand side of (\ref{Theorem}), 
while under most circumstances negligible, 
may be profitably used to obtain upper bounds on 
the possible condensate fraction as a function of 
temperature\cite{TonyCondFract}.

To demonstrate that the inequality (\ref{RcCF})
is indeed useful, I work out an
explicit example for an experimentally easy-to-access case. 
That is, I consider a Thomas-Fermi condensate wave function in 
a quasi-1D harmonic trap with $\omega_z \ll \omega_\perp$, which reads  
$\Phi_0 (z) = ({n^0_{\rm TF}} 
\left(1-{z^2}/{Z_{\rm TF}^2} \right)/N)^{1/2}$. It is 
assumed that $F$ is sufficiently close to unity, $N_0 \simeq N$, 
which amounts in effect to taking the limit of both sides of (\ref{RcCF}) 
to linear order in $1-F$.  
The resulting function ${\cal C}$ is strongly peaked at its global maximum 
$k_m \simeq 3.7 Z_{TF}^{-1}$ ($\lambda_m \simeq 1.7 Z_{\rm TF})$,  
where ${\cal C}(k_m)\simeq 1.54\times 10^{-2}$. It should be noted 
that, if one were to make the simple ``guess'' of, say, 
$\lambda_m = Z_{\rm TF}$, not using a variational ansatz for the 
right-hand side of (\ref{RcCF}), one would obtain a 
(much) weaker bound, and would therefore strongly limit the 
usefulness of the inequality. 

Using (\ref{Rcdef}), 
we have $R_c = 4Z_{\rm TF}/3\sqrt{2\pi}$, which tell us that, 
as expected, the effective radius of curvature is of order 
the Thomas-Fermi length of the strongly elongated condensate. 
Neglecting the second term on the right-hand side of (\ref{RcCF}), 
at $k=k_m$ the rigorous inequality    
$
{Z_{\rm TF}}/{\lambda_{\rm dB}} \le 6 \times 
\, \sqrt{{N-N_0}} 
$ 
is obtained. This result should be compared to the square root of 
relation (\ref{1D2DDivergence}), which reads, at $n_0 \rightarrow n$, 
$R/\lambda_{\rm dB} \le (k_{\rm Pl}/n) \sqrt{N-N_0}$. 
The bound thus obtained on the maximal Thomas-Fermi length of a thin 
BEC cigar is  consistent with experiments\cite{Ketterle1D2D,HT}. 
It is important to stress, however, that this
would not be the case, had we neglected, following the 
Bogoliubov approximation, the second term under 
the  square in (\ref{Ckdef}). Doing so reduces the right-hand side
of the above inequality by about an order of magnitude for 
quasi-1D Thomas-Fermi condensates, and 
leads to a contradiction with experiment.  
Therefore, in this sense, particle number conservation, 
violated by the standard Bogoliubov prescription, 
is necessary for the condensate to exist.

The conditions needed to violate the Thomas-Fermi relation on 
$Z_{\rm TF}$ 
become transparent in experimentally accessible parameters 
if we write 
$Z_{\rm TF}= (6N d_z^2 a_s /|1-C a_s/d_\perp|)^{1/3}
({\omega_\perp}/{\omega_z})^{1/3} $, where
$C=1.4603$, 
a relation valid as long as the 3D scattering length 
is much less than the oscillator length in the perpendicular 
direction, $a_s\ll d_\perp$\cite{Dunjko}. 
Then 
\begin{equation}
\frac{\omega_\perp}{\omega_z}
\le 36.7\, 
\frac{\lambda^3_{\rm dB}}{d_z^2 a_s}
\left| 1- C \frac{a_s}{d_\perp} \right| N^{1/2} f^3,  
\label{aspectratio}
\end{equation}
where $f=0$ if all particles are in the condensate, and assumes
its maximal value $f=1$ if there are none.  Though the inequality in 
the above form is strictly valid only if $f\ll 1 $, for a first estimate 
of the critical aspect ratio it is sufficient to let $f=1$.  
From relation (\ref{aspectratio}) it is apparent that, 
leaving $\omega_z\,(d_z)$, $a_s$ and $N$ fixed, 
the inequality is most easily violated (apart
from increasing the temperature), if the aspect 
ratio is increased  by increasing the transverse trapping 
${\omega_\perp}$. 
This route of increasing the aspect ratio 
to check relation (\ref{aspectratio}) 
should in particular be possible in the narrow condensate 
tubes created in 2D optical lattices\cite{Moritz}, 
where the aspect ratios can be increased to 1000 and beyond.
Indeed, in a such a 2D optical lattice, a transition from a 
superfluid to a Mott insulating phase with no long-range order present  
has been observed, where the point of transition should 
fulfill inequality (\ref{aspectratio})\cite{Stoeferle}.
Also note that close to the geometric scattering resonance\cite{Dunjko} at 
$a_s= d_\perp/C$,  
the right-hand side of (\ref{aspectratio}) approaches zero and the critical 
aspect ratio strongly decreases, which may be used to detect the 
scattering resonance. 

To conclude, I stress that the particular advantage of 
relation (\ref{RcCF}) is its independence on specific model  
Hamiltonians for the strongly anisotropically confined gas. Bounds for 
the existence of condensates can thus also be given where explicit 
results are not available, e.g., for long-range interaction potentials.  
By its very nature, as an inequality, it does not (and cannot) {\em prove}   
that a certain condensate exists. For such a proof one still has to resort 
to an explicit (many-body) calculation, to show 
that the state with single-particle 
wave function $\Phi_0$ is macroscopically occupied. 
However, on the other hand, 
the inequality can rule out parameter ranges for which any given  set
$\{N_0, \Phi_0({\bm r}), n({\bm r}) \}$, at a specified finite temperature, 
is {\em not} possible. 
Essentially, it answers the following concrete question: 
How much can I stretch a
strongly elongated, trapped and inhomogeneous 
Bose-Einstein condensate until its long-range coherence breaks?


\end{document}